\documentclass[final, letterpaper]{IEEEtran}

\usepackage[margin=1.32cm, centering]{geometry}
\usepackage{tikz}
\usepackage[sumlimits,intlimits]{amsmath}
\usepackage{amsfonts,amssymb}%
\usepackage{bm}%
\usepackage{graphicx,graphics}%
\usepackage[noadjust]{cite}%
\usepackage{color}%
\usepackage{url}
\usepackage{verbatim}
\usepackage{balance}
\usepackage{multirow}
\usepackage{multicol}
\usepackage{stfloats}
\usepackage{xspace}
\usepackage{enumerate}
\usepackage{rotating}
\usepackage[tight]{subfigure}
\usepackage{algorithm}
\usepackage{algorithmicx, algpseudocode}
\usepackage{epstopdf}

\newtheorem{lemma}{Lemma}

\newcommand{\siga}{\sigma^2_{a}}
\newcommand{\sigc}{\sigma^2_{c}}

\newcommand{\ds}{\displaystyle}
\newcommand{\la}{\langle}
\newcommand{\ra}{\rangle}
\newcommand{\clN}{{\cal N}}
\newcommand{\Tr}{{\sf Trace}}
\ifCLASSOPTIONpeerreview

\fi

\begin{document}

\title{Path-Following Algorithms for Beamforming and Signal Splitting in RF Energy Harvesting Networks}

%

\author{Ali~A.~Nasir, Hoang~D.~Tuan, Duy~T.~Ngo, Salman~Durrani, and Dong In Kim%
\thanks{Manuscript received 16 March 2016, revised 12 May 2016 and accepted 7 June 2016. The associated editor handling the review of this letter is W.~K.~Ng.}
\thanks{A.~A. Nasir is with the National University of Sciences and Technology, Pakistan. H.~D. Tuan is with University of Technology Sydney, Australia. D.~T. Ngo, the corresponding author, is with The University of Newcastle, Australia (email: duy.ngo@newcastle.edu.au).  S. Durrani is with the Australian National University. D.~I. Kim is with Sungkyunkwan University, South Korea. This work was supported in part by the National Research Foundation of Korea (NRF) grant funded by the Korean government (MSIP) (2014R1A5A1011478).}
}

\maketitle

\begin{abstract}
We consider the joint design of transmit beamforming and receive signal-splitting ratios in the downlink of
a wireless network with simultaneous radio-frequency (RF) information and energy transfer. Under constraints on the signal-to-interference-plus-noise ratio (SINR) at each user and the total transmit power at the base station, the design objective is to maximize either the sum harvested energy or the minimum harvested energy. We develop a computationally efficient path-following method to solve these challenging nonconvex optimization problems. We mathematically show that the proposed algorithms iteratively progress and converge to locally optimal solutions. Simulation results further show that these locally optimal solutions are the same as the globally optimal solutions for the considered practical network settings.
\end{abstract}


\section{Introduction}\label{sec:int}
RF information and energy transfer, which treats wireless receivers as either conventional information decoding (ID) receivers or energy harvesting (EH) receivers, has emerged as an attractive paradigm for green communication in the next-generation wireless networks \cite{Lu-14-A,Chen-15-Apr-A}. In particular, transmit beamforming to improve the quality-of-service of  RF
information and energy transfer has drawn significant research interest \cite{Tietal15,Letal15}.
There are two important beamforming problems in optimizing the  harvested energy: (i) maximization of the total harvested energy (i.e., `sum EH maximization problem'), and (ii) maximization of the most disadvantaged EH receiver in the network (i.e., `max-min EH problem'). Both problems are subject to the minimum SINR constraint at the ID receivers and the total transmit power constraint at the base station (BS). They are indefinite quadratic programs in the beamforming vectors, which
are typically recast as nonconvex rank-one matrix constrained semidefinite programs (SDPs) in the beamforming outer products. The rank-one matrix constraints are then omitted, allowing for suboptimal SDP relaxation \cite{Boyd-04-B}.


Meanwhile, it is realistic for the users located in the vicinity of the BS to conduct both ID and EH functions
by a signal splitting (SS) based receiver \cite{Nasir-13-A,Lu-14-A,Kretal14,Letal15, Vu-15-A, Ng-14-A}. A problem of particular interest is how to jointly design the transmit beamforming vectors and the receive SS ratios in order to maximize the harvested energy. Finding an efficient computational solution for this problem is a major challenge in itself. Due to the strong coupling of the beamforming vectors and the SS ratios in the optimization
objective, the sum EH maximization problem
cannot be recast as a matrix rank-one constrained SDP to accommodate the conventional SDP relaxation approach \cite{Ng-14-A}. For the max-min EH problem, the beamforming vectors and SS ratios can be decoupled via a bisection search for the worst EH receiver. Nevertheless, the resultant SDP relaxation only provides an upper bound performance for this problem as this approach is less likely to generate a rank-one matrix solution. Randomization in conjunction with linear programming must be further employed to generate the beamforming vectors, which however could be far away from the actual optimal solutions \cite{Zhang-13-Oct-P}.

This letter aims to develop an efficient computational method for the two aforementioned joint design EH optimization problems. Our research contributions are summarized as follows.
\begin{itemize}
\item We propose a new path-following method for their solutions. Each iteration of the proposed algorithms requires solving a second-order cone program (SOCP) in the beamforming vectors and SS ratios. We mathematically show that the algorithms progress at every iteration and converge to locally optimal solutions.  In addition, our simulation results with practical parameter values show that the obtained locally optimal solutions are the same as the upper bound given by the exhaustive search over the domain of SS ratios (for the sum EH maximization problem) or that by the SDP relaxation-based bisection search (for the max-min EH problem). This demonstrates the usefulness of our solutions.

\item Our algorithms are computationally efficient and simple to implement. {\color{black}In contrast,} the exhaustive search and the SDP relaxation-based bisection search can only achieve the upper bounds after solving many SDPs in the beamforming outer products of a substantially increased dimension. Such techniques are computationally prohibitive in practical networks.
\end{itemize}

\section{System Model and Problem Formulation}\label{sec:sys_mod}

We consider an energy-constrained small-cell wireless network where a BS with $M>1$ antennas transmits to $N$ single-antenna users (UEs).
Let $\mathbf{h}_{n} \in \mathbb{C}^{M \times 1}$ be the flat fading channel vector between the BS and UE $n  \in \mathcal{N} \triangleq \{1,\dots,N\}$, which includes the effects of large-scale pathloss and small-scale fading.
Let $x_{n}$ (with $\mathbb{E}\{ | x_{n} |^2 \} = 1$) denote the intended message for  UE $n$, which
is beamformed by the vector $\mathbf{w}_{n} \in \mathbb{C}^{M \times 1}$ at the BS.
The baseband signal received by UE $n$ is expressed as:
\begin{align}\label{eq:y_kn_explicit}
 y_{n} &= \mathbf{h}_{n}^H \mathbf{w}_{n} x_{n} + \mathbf{h}_{n}^H  \sum_{\eta\in\mathcal{N}\setminus{\{n\}}} \mathbf{w}_{\eta} x_{\eta}    +  z_{n}^a,
\end{align}
where  $z_{n}^a$ is the zero-mean circularly symmetric additive white Gaussian noise (AWGN) with variance $\siga$ introduced by the receive antenna. We assume perfect channel state information (CSI) is available at the BS, where details of the CSI acquisition process can be found in \cite{Xu-14-Sep-A}. The first term in \eqref{eq:y_kn_explicit} is the intended signal for UE $n$ while the second term represents the interference.

Given that RF EH is only practical when the BS-UE distance is sufficiently small \cite{Lu-14-A}, our model divides the BS coverage area into (i) the EH zone (near the BS), and (ii) the ID zone (outside the EH zone). If a UE resides inside the EH zone, it can conduct both EH and ID; otherwise, it only performs ID. Assume there are $N_1$ and $N_2$ UEs residing in the EH and ID zones, respectively, where $N_1 + N_2 = N$. Let us index the EH-ID UEs by $n_1 \in \mathcal{N}_{1} \triangleq \{1,\dots,N_{1}\}$ and the ID-only UEs by $n_2 \in \mathcal{N}_{2} \triangleq \{N_1+1,\dots,N\}$.

At an EH-ID UE $n_1\in\mathcal{N}_1$, the signal splitter divides the received signal $y_{n_1}$ into two parts in the proportion of $\alpha_{n_1}^2:(1-\alpha_{n_1}^2)$, where $\alpha_{n_1}^2 \in (0,1)$ is termed as the SS ratio for UE $n_1$. The first part $ \alpha_{n_1} y_{n_1}$ forms an input to the ID receiver as:
\begin{align}\label{eq:first_part}
   \alpha_{n_1} y_{n_1} + z_{n_1}^c &= \alpha_{n_1} \bigg(   \mathbf{h}_{n_1}^H \sum_{\eta\in\mathcal{N}} \mathbf{w}_{\eta} x_{\eta}  +  z_{n_1}^a \bigg) + z_{n_1}^c, \vspace{-1cm}
\end{align}
where $z_{n_1}^c \sim  \mathcal{CN}(0,\sigc)$ is the additional noise  with variance $\sigc$ introduced by the ID receiver circuitry. From \eqref{eq:first_part}, we write the general expression of the SINR at the input of the ID receiver of UE $n \in \mathcal{N}$ (i.e., either an EH-ID UE or an ID-only UE) as:
\begin{equation}\label{sinr}
    \text{SINR-UE}_{n}\triangleq |\mathbf{h}_{n}^H \mathbf{w}_{n} |^2/\varphi_{n}(\mathbf{w},\alpha_n),
\end{equation}
    where $\mathbf{w} \triangleq [\mathbf{w}_{n}]_{n\in {\cal N}}$ and $\varphi_{n}(\mathbf{w},\alpha_n)\triangleq \sum_{\eta\in\mathcal{N}  \setminus\{n\}} |\mathbf{h}_{n}^H \mathbf{w}_{\eta} |^2
 + \siga + \sigc/\alpha_{n}^2$ with $\alpha_n\in (0,1)$
 for $n\in \mathcal{N}_1$ and $\alpha_n = 1$ for $n \in \mathcal{N}_2 $.

Also at the EH-ID UE $n_1\in\mathcal{N}_1$, the second part $\sqrt{1-\alpha_{n_1}^2} y_{n_1}$ of the received signal $y_{n_1}$ is processed by an EH receiver. The energy harvested by UE $n_1$ is thus given by:
\begin{align}\label{harvest}
   E_{n_1}(\mathbf{w},\alpha_{n_1}) &\triangleq \zeta_{n_1}(1-\alpha_{n_1}^2)p_{n_1}(\mathbf{w}) ,
\end{align}
where $p_{n_1}(\mathbf{w})\triangleq    \sum_{\eta\in\mathcal{N}} | \mathbf{h}_{n_1}^H \mathbf{w}_{\eta}  |^2  + \siga$,
and the constant $\zeta_{n_1} \in (0,1)$ denotes the efficiency of energy conversion at the EH receiver. Upon defining
$\boldsymbol\alpha=[\alpha_{n_1}]_{n_1\in\mathcal{N}_1}$, the sum EH maximization problem is formulated as:
\begin{subequations} \label{eq:P1}
\begin{align}
    \ds\underset{\substack{\mathbf{w}_n \in \mathbb{C}^{M\times 1}, \forall  n \in \mathcal{N}, \\ \boldsymbol\alpha\in(0,1)^{N_1} }}{\max} &F(\mathbf{w},\boldsymbol\alpha)\triangleq
    \ds\sum_{n_1\in\mathcal{N}_1}\zeta_{n_1}(1-\alpha_{n_1}^2)p_{n_1}(\mathbf{w})
            \label{eq:O1}\\
         \text{s.t.}\quad  &\ds\sum_{n\in\mathcal{N}  } \| \mathbf{w}_{n} \|^2\leq P,\label{eq:01a}\\
&\ds |\mathbf{h}_{n}^H \mathbf{w}_{n} |^2\geq \gamma_{n}^{\min}\varphi_{n}(\mathbf{w},\alpha_n), \  n\in\mathcal{N}.   \label{eq:C1}
\end{align}
\end{subequations}
Constraint \eqref{eq:01a} caps the total transmit power at a predefined value $P$, and constraint \eqref{eq:C1} ensures that the received SINR by UE $n$ be greater than a predefined threshold $\gamma_{n}^{\min}$. Note that \eqref{eq:P1} is a nonconvex optimization problem
because \eqref{eq:O1} is not concave.

In the relaxation approach, by omitting the difficult constraints $\mbox{rank}(\mathbf{W}_n)=1$
for the beamforming outer products $\mathbf{W}_n=\mathbf{w}_n\mathbf{w}_n^H\in\mathbb{C}^{M\times M}$,
 one uses  $|\mathbf{h}_{n}^H \mathbf{w}_{n} |^2=\mathbf{h}_{n}^H \mathbf{W}_{n}\mathbf{h}_{n}$, $| \mathbf{h}_{n_1}^H \mathbf{w}_{n}  |^2=
\mathbf{h}_{n_1}^H \mathbf{W}_{n}\mathbf{h}_{n_1}$ and $\| \mathbf{w}_{n} \|^2=\Tr(\mathbf{W}_{n})$  to form
\begin{subequations} \label{SDP1}
\begin{align}
    \ds\max_{\overset{ \mathbf{W}_n \in \mathbb{C}^{M\times M}, \forall  n \in \mathcal{N},}{\boldsymbol\alpha\in(0,1)^{N_1}}}&
    \ds\sum_{n_1\in\mathcal{N}_1}\zeta_{n_1}(1-\alpha_{n_1}^2)\tilde{p}_{n_1}(\mathbf{W})\label{SDP1a}\\
     \text{s.t.}& \ds\sum_{n\in\mathcal{N}  } \Tr(\mathbf{W}_{n}) \leq P, \ \mathbf{W}_n\succeq \mathbf{0}, \, n\in \clN,   \label{SDP1b}\\
&\ds \mathbf{h}_{n}^H \mathbf{W}_{n}\mathbf{h}_{n}\geq \gamma_{n}^{\min}\tilde{\varphi}_{n}(\mathbf{W},\alpha_n), \  n\in\mathcal{N}, \label{SDP1c}
\end{align}
\end{subequations}
with linear function  $\tilde{p}_{n_1}(\mathbf{W})=\sum_{\eta\in\mathcal{N}}\mathbf{h}_{n_1}^H \mathbf{W}_{\eta}\mathbf{h}_{n_1} + \siga$ and convex function $\tilde{\varphi}_n(\mathbf{W},\alpha_n)=\sum_{\eta\in\mathcal{N}  \setminus\{n\}} \mathbf{h}_{n}^H \mathbf{W}_{\eta}\mathbf{h}_{n}
 + \siga + \sigc/\alpha_{n}^2$, where $\mathbf{W}\triangleq [\mathbf{W}_\eta]_{\eta\in\clN}$. The multiplicative
objective function in (\ref{SDP1a}) remains nonlinear nonconcave in $\mathbf{W}$ and
$\boldsymbol\alpha$, making (\ref{SDP1}) still computationally difficult. In the following section, we propose an efficient path-following method to solve \eqref{eq:P1}
directly in the beamforming vectors $\mathbf{w}_n\in\mathbb{C}^M$ and the scalar SS ratios $\alpha_{n_1}$.


\section{SOCP-based Iterative Optimization}
Since the power constraint (\ref{eq:01a}) is convex quadratic, let us first deal with the SINR constraint \eqref{eq:C1}. For $\bar{\mathbf{w}}_{n}=e^{-\jmath.{\sf arg}(\mathbf{h}_{n}^H
\mathbf{w}_{n})}\mathbf{w}_{n}$, one has $|\mathbf{h}_{n}^H
\mathbf{w}_{n}|=\mathbf{h}_{n}^H\bar{\mathbf{w}}_{n}={\sf Re}\{\mathbf{h}_{n}^H
\bar{\mathbf{w}}_{n}\}\geq 0$ and $|\mathbf{h}_{n'}^H
\mathbf{w}_{n}|=|\mathbf{h}_{n'}^H\bar{\mathbf{w}}_{n}|$ for $n'\neq n$, where $\jmath \triangleq \sqrt{-1}$ and ${\sf Re}\{x\}$ is the real part of a complex number $x$.
Therefore, \eqref{eq:C1} can be recast as
$\ds {\sf Re}\{\mathbf{h}_{n}^H \mathbf{w}_{n}\} \geq \sqrt{\gamma_{n}^{\min}\varphi_{n}(\mathbf{w},\mathbf{\alpha}_{n})}, \  n\in\mathcal{N}$ \cite{WLP06}. The latter is equivalent to the following SOC:
\begin{align}
 \ds {\sf Re}\{\mathbf{h}_{n}^H \mathbf{w}_{n}\} &\geq \sqrt{\gamma_n^{\min}}
\left\| \begin{matrix}\sigma_a\cr
\mu_n \cr
\left(\mathbf{h}_n^H\mathbf{w}_{{\color{black} \eta}}\right)_{{\color{black} \eta}\in \mathcal{N}\setminus \{n\}}
\end{matrix}
\right\|_{2},  \ n\in\mathcal{N}, \label{soc1} \\
\left(\begin{matrix}t_{n_1}&1\cr 1&\alpha_{n_1}\end{matrix}\right)&\succeq {\color{black} \mathbf{0} }, \  n_1\in\mathcal{N}_1,
\label{soc2}
\end{align}
where $\left(\mathbf{h}_n^H\mathbf{w}_{{\color{black} \eta}}\right)_{{\color{black} \eta}\in \mathcal{N}\setminus \{n\}}$ is an $(N-1) \times 1$ column vector, $t_n$ is an auxiliary variable, and
$\mu_n = \sigma_{c} t_{n} $ for $n \in \mathcal{N}_1$ whereas $\mu_n= \sigma_{c}$ for $n \in \mathcal{N}_2$. This means
(\ref{eq:P1}) is a nonconcave maximization problem subject to convex constraints. To develop a path-following procedure
for the computational solution of (\ref{eq:P1}), the following lemma provides an effective concave lower approximation for the
multiplicative objective in \eqref{eq:O1} \cite{Marks-78-A}.
\begin{lemma}\label{prop} {\it For any $\mathbf{w}$, $\mathbf{w}^{(\kappa)}$ and $\alpha_{n_1}\in (0,1)$,
$\alpha_{n_1}^{(\kappa)}\in (0,1)$ the following relations hold true:
\begin{subequations} \label{eq:P_max_min11}
\begin{align}
(1-\alpha_{n_1}^2)p_{n_1}(\mathbf{w}) &\ge {\color{black}\breve{p}}_{n_1}^{(\kappa)}(\mathbf{w},\alpha_{n_1}),\label{in1}\\
(1-(\alpha_{n_1}^{(\kappa)})^2)p_{n_1}(\mathbf{w}^{(\kappa)})&= {\color{black}\breve{p}}_{n_1}^{(\kappa)}(\mathbf{w}^{(\kappa)},\alpha_{n_1}^{(\kappa)})
\label{in2}
\end{align}
 \end{subequations}
for ${\color{black}\breve{p}}_{n_1}^{(\kappa)}(\mathbf{w},\alpha_{n_1}) \triangleq $
\begin{equation}\label{pk}
\begin{array}{r}
 \ds2(1-(\alpha_{n_1}^{(\kappa)})^2)\big[\ds  \sum_{{\color{black} \eta}\in\mathcal{N}}
        \Re\{(\mathbf{w}_{{\color{black} \eta}}^{(\kappa)})^H\mathbf{h}_{n_1}\mathbf{h}^H_{n_1}
        \mathbf{w}_{{\color{black} \eta}} \}+\sigma_a^2\big] \\ -\ds\frac{p_{n_1}(\mathbf{w}^{(\kappa)})(1-(\alpha_{n_1}^{(\kappa)})^2)^2}{1-\alpha_{n_1}^2},
\end{array}
\end{equation}
which is concave in $(\mathbf{w},\alpha_{n_1})$.
}
\end{lemma}
\begin{IEEEproof}
See the Appendix.
\end{IEEEproof}

With Lemma \ref{prop}, we propose Algorithm \ref{alg:1} to solve problem (\ref{eq:P1}).
\begin{algorithm}[!t]\caption{SOCP-Based Iterative Optimization for Sum EH Maximization Problem \eqref{eq:P1} }\label{alg:1}
  \begin{algorithmic}[1]
  \State Initialize $\kappa := 0$.
  \State Find a feasible point $(\mathbf{w}^{(0)}, \boldsymbol\alpha^{(0)})$ of \eqref{eq:P1} by solving the SOCP:
  $\ds\min_{ \substack{\mathbf{w}_{n},t_{n_1},\boldsymbol\alpha\in(0,1)^{N_1}, \\ \ \forall  n \in \mathcal{N}, n_1 \in \mathcal{N}_1}} \sum_{n\in\mathcal{N}} \| \mathbf{w}_{n} \|^2$ s.t.  (\ref{soc1}), (\ref{soc2}).
  \Repeat
  \State Solve the SOCP:
  \begin{align}\label{itea}
 \ds\max_{\overset{\mathbf{w}_{n},t_{n_1} , {\color{black}\forall  n \in \mathcal{N}, n_1 \in \mathcal{N}_1}}{\boldsymbol\alpha\in(0,1)^{N_1}}}\ & F^{(\kappa)}(\mathbf{w},\boldsymbol\alpha)\triangleq
 \sum_{n_1\in\mathcal{N}_1}\zeta_{n_1} {\color{black}\breve{p}}_{n_1}^{(\kappa)}(\mathbf{w},\alpha_{n_1}) \notag \\
      &  \hspace{0cm}  \ \text{s.t.} \ \eqref{eq:01a}, \eqref{soc1}, \eqref{soc2}.
\end{align}
   \quad \ \ to generate a feasible point $(\mathbf{w}^{(\kappa+1)},\boldsymbol\alpha^{(\kappa+1)})$ for \eqref{eq:P1}.
  \State Set $\kappa := \kappa+1$.
  \Until{convergence of the objective in \eqref{eq:O1}.}
  \end{algorithmic}
\end{algorithm}
Note that $F(\mathbf{w}^{(\kappa+1)},\boldsymbol\alpha^{(\kappa+1)})\geq F^{(\kappa)}(\mathbf{w}^{(\kappa+1)},\boldsymbol\alpha^{(\kappa+1)})$ by (\ref{in1}) and $F(\mathbf{w}^{(\kappa)},\boldsymbol\alpha^{(\kappa)})= F^{(\kappa)}(\mathbf{w}^{(\kappa)},\boldsymbol\alpha^{(\kappa)})$ by (\ref{in2}), while
$F^{(\kappa)}(\mathbf{w}^{(\kappa+1)},\boldsymbol\alpha^{(\kappa+1)})>
F^{(\kappa)}(\mathbf{w}^{(\kappa)},\boldsymbol\alpha^{(\kappa)})$ as long as $(\mathbf{w}^{(\kappa+1)},\boldsymbol\alpha^{(\kappa+1)})\neq (\mathbf{w}^{(\kappa)},\boldsymbol\alpha^{(\kappa)})$
because $(\mathbf{w}^{(\kappa)},\boldsymbol\alpha^{(\kappa)})$ is also feasible to (\ref{itea}). As a result,
\[
F(\mathbf{w}^{(\kappa+1)},\boldsymbol\alpha^{(\kappa+1)})>F^{(\kappa)}(\mathbf{w}^{(\kappa)},\boldsymbol\alpha^{(\kappa)}),\]
i.e., $(\mathbf{w}^{(\kappa+1)},\boldsymbol\alpha^{(\kappa+1)})$ is a better feasible point to (\ref{eq:P1}) than
$(\mathbf{w}^{(\kappa)},\boldsymbol\alpha^{(\kappa)})$. The sequence $\{(\mathbf{w}^{(\kappa)},\boldsymbol\alpha^{(\kappa)})\}$ of improved feasible points
to (\ref{eq:P1}) converges at least to its local optimum, which also satisfies the Karush-Kuhn-Tucker condition \cite{Marks-78-A}. {\color{black}Simulation results in Sec.~\ref{sec:simulation} further show that our algorithm achieves the computable upper bound, implying that a global optimum is attained in the practical settings considered in the simulations.}

Our proposed method can be extended to solve the following max-min EH problem:
\begin{equation} \label{eq:P2}
    \ds\max_{{\color{black}\mathbf{w}},\boldsymbol\alpha\in(0,1)^{N_1}}\
    \ds\min_{n_1\in\mathcal{N}_1}\zeta_{n_1}(1-\alpha_{n_1}^2)p_{n_1}(\mathbf{w})\ \ \mbox{s.t.}
            \ \ \eqref{eq:01a}, \eqref{eq:C1}.
\end{equation}
In this case, instead of (\ref{itea}), we solve the following SOCP to generate $(\mathbf{w}^{(\kappa+1)}, \boldsymbol\alpha^{(\kappa+1)})$ from
$(\mathbf{w}^{(\kappa)}, \boldsymbol\alpha^{(\kappa)})$:
\begin{equation}\label{iteap2}
 \max_{\overset{{\color{black}\mathbf{w},t_{n_1}, \forall n_1}}{\boldsymbol\alpha\in(0,1)^{N_1}}}\ \ds
 \min_{n_1\in\mathcal{N}_1} \zeta_{n_1}{\color{black}\breve{p}}_{n_1}^{(\kappa)}(\mathbf{w},\alpha_{n_1})
          \ \ \text{s.t.} \ \ \eqref{eq:01a}, \eqref{soc1}, \eqref{soc2}.
\end{equation}
{\bf Remark.} {\color{black}Unlike {\color{black}\cite{Zhang-13-Oct-P,Xu-14-Sep-A}}, here we consider both EH and ID functionalities for the near UEs. The nature of the optimization problems is thus nontrivially changed, requiring a different solution approach as has been proposed in this letter.} {\color{black} The difference-of-two-convex-functions based optimization approach in \cite{Vu-15-A} is not suitable for the sum EH maximization problem \eqref{eq:P1}. Not only would the problem dimension be increased, it is also not easy to find a good feasible initial point to use in this approach due to the nonconvex constraints. Recently, \cite{Shi-14-Dec-A} has employed the SOCP (\ref{soc1})-(\ref{soc2}) to express the SINR constraint (\ref{eq:C1}) in the problem of minimizing the total transmit power $\sum_{n\in{\cal N}}||\mathbf{w}_n||^2$. The EH threshold is then set to make the EH constraints automatically satisfied by the SINR constraints
(\ref{soc1})-(\ref{soc2}). The approach in \cite{Shi-14-Dec-A} is thus inapplicable to our EH optimization problems (\ref{eq:P1}) and (\ref{eq:P2}).
}

\vspace{-0.25cm}
\section{Numerical Examples}\label{sec:simulation}
We consider a small-cell network with $N=6$ UEs and $M = \{6,7,8\}$ antennas at the BS. {\color{black} Unless specified otherwise}, the BS-to-UE distance is set as $7$ m and $20$ m for the $N_1=3$ EH-ID UEs and the $N_2=3$ ID-only UEs, respectively. {\color{black}Later in Fig. \ref{fig:comp}, we will consider both $7$ m and $9$ m distances for the $N_1$ EH-ID UEs since commercial radiative wireless charging systems (e.g., the Cota systems \cite{cota}) can deliver power up to $30$ ft (i.e., $9$~m).} {\color{black}We assume a simplified path loss model with  carrier  center  frequency  of $470$ MHz, transmit antenna gain of $10$ dBi, reference distance of $2$ m and path loss exponent of $2.6$ \cite{Ng-14-A}, \cite[Sec. 2.6]{Goldsmith-05-B}.}
 We generate a Rician fading channel with a Rician factor of $K_R=10$ dB.  For simplicity and without loss of generality, we assume that $\zeta_{n} = \zeta$ and $\gamma_n^\text{min} = \gamma^\text{min}, \ \forall n\in\mathcal{N}$. Thus, we set $\zeta = 0.5$, $\siga = -90$ dBm, and $\sigc = -90$ dBm. The numerical results are averaged over $1,000$ random channel realizations.

\subsection{Results for Sum EH Maximization Problem  \eqref{eq:P1}}

\begin{figure}[t]
    \centering
    \includegraphics[width=0.4 \textwidth]{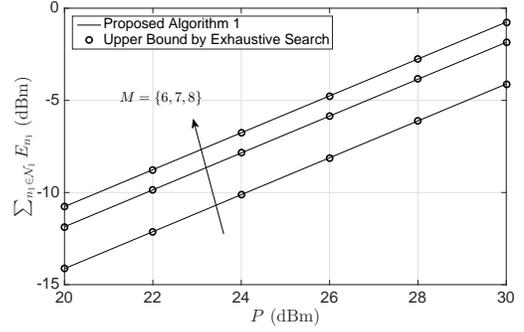}
  \caption{Optimized sum harvested energy for $\gamma^\text{min} = 12$ dB.}
  \label{fig:pow}
\end{figure}

Fig.~\ref{fig:pow} plots the optimized sum harvested energy for $P = \{20,22,\hdots,30\}$ dBm and $\gamma^\text{min} = 12$ dB.  On average, Algorithm 1 requires only $6.5$ iterations (i.e., solving $6.5$ SOCPs in $\alpha_{n_1}$ and $\mathbf{w}_n$). Fig.~\ref{fig:pow} also confirms that the sum harvested energy increases when more transmit power is available. For the typical value of $P = 26$ dBm used in a small-cell BS, the sum harvested energy for the three EH-ID UEs is found as {\color{black}$\{-8.1,-5.8,-4.7\}$ dBm} for $M = \{6,7,8\}$ BS antennas.

\begin{figure}[t]
    \centering
    \includegraphics[width=0.4 \textwidth]{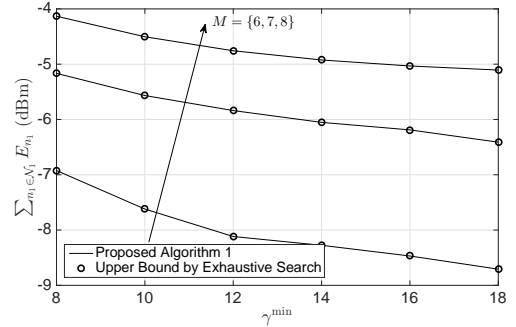}
  \caption{Optimized sum harvested energy for $P = 26$ dBm.}
  \label{fig:gam}
\end{figure}

Fig.~\ref{fig:gam} plots the sum harvested energy for $\gamma^\text{min} = \{8,10,\hdots,18\}$ dB and $P = 26$ dBm. As can be seen from the figure, increasing the SINR threshold reduces the harvested amount of energy. This is because more received power is diverted to the ID receiver to meet the minimum SINR requirement, leaving less power for the EH receiver. Fig.~\ref{fig:gam} shows that the sum harvested energy is in the range of {\color{black}$-9$ dBm to $-4$ dBm} for the practical network parameters considered.

\textit{Convergence and Complexity:} Figs.~\ref{fig:pow} and \ref{fig:gam} demonstrate that Algorithm 1 practically yields the globally optimal solution of
(\ref{eq:P1}). {\color{black}It achieves the upper bound obtained by a branch-and-bound (BB) procedure \cite{Tuy-B-98}
over the domain $(0,1)^{N_1}$ that finds the optimal SS ratios $[\alpha_{n_1}]_{n_1\in\clN_1}$ of the nonconvex program (\ref{SDP1}). At each iteration of this procedure, an upper bound of the nonconvex program (\ref{SDP1}) over $[p,q]\subset (0,1)^{N_1}$ is provided by the SDP $\ds\max_{\mathbf{W},\alpha_{n_1}\in[p_{n_1},q_{n_1}],\forall n_1}\
    \ds\sum_{n_1\in\mathcal{N}_1}\zeta_{n_1}(1-\alpha_{n_1}^2)\tilde{p}_{n_1}(\mathbf{W})\
     \text{s.t.}\ \eqref{SDP1b}-\eqref{SDP1c}$. The incumbent SS ratios are updated by solving (\ref{SDP1})  at a fixed $\alpha_{n_1}=(p_{n_1}+q_{n_1})/2$,
$n_1\in\clN_1$,     which is an SDP in the beamforming outer products $\mathbf{W}_n$.
      While much more efficient than the intensive grinding, the computational complexity of this exhaustive search over the domain $(0,1)^{N_1}$ is still prohibitively high. } Furthermore, the total dimension of the beamforming outer products $\mathbf{W}_n$ is $NM(M+1)/2\in \{126, 168, 216\}$ compared with
$NM\in\{36,42,48\}$ of the beamforming vectors $\mathbf{w}_n$. In our simulations, we have also observed that {\color{black}this search} gives solution {\color{black}$\mathbf{W}_n$} with rank greater than one in $50\%$ and almost $100\%$ of all cases for $\gamma^{\min}=12$ dB and $\gamma^{\min}=18$ dB, respectively.

\subsection{Results for Max-Min EH Problem (\ref{eq:P2})}
\begin{figure}[t]
    \centering
    \vspace{-2pt}
    \includegraphics[width=0.4 \textwidth]{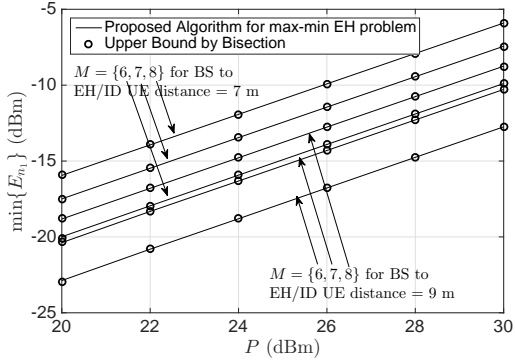}
  \caption{Maximized minimum UE harvested energy for $\gamma^\text{min} = 12$ dB.}
    \label{fig:comp}
    \vspace{-2pt}
\end{figure}

{\color{black} Fig.~\ref{fig:comp} plots the optimized energy harvested by the worst user over a range of transmit power budget $P = \{20,22,\hdots,30\}$ dBm. Here, the distances from the BS to all EH-ID UEs are set as $7$ m and $9$ m.} It is observed that our algorithm achieves
the upper bound given by the SDP-based bisection search with considerably less computational complexity. The bisection search locates the largest $\lambda$ such that the SDP
{\color{black}$\zeta_{n_1} \tilde{p}_{n_1}(\mathbf{W}) \ge \lambda/(1-\alpha_{n_1}^2), n_1\in \clN_1,
\eqref{SDP1b}-\eqref{SDP1c}$ is feasible in the beamforming outer products $\mathbf{W}_n$ and
the scalar SS ratios $\alpha_{n_1}$.}  On average, our algorithm converges after $6.8$ iterations (i.e., solving $6.8$ SOCPs) whereas the bisection search solves $11.6$ SDPs  of a much larger dimension. Furthermore, the latter often yields a rank-greater-than-one matrix $\mathbf{W}_n$. Particularly, for $M = 8$ BS antennas, we have observed $\mathbf{W}_n$ with rank greater than one in $90\%$ of the simulation cases. To generate a rank-one matrix for extracting beamforming vectors, a randomization step in conjugation with linear programming is required in such approach. {\color{black} This incurs extra computational overhead} while the performance may suffer \cite[Fig. 3]{Zhang-13-Oct-P} as the extracted point may be far away from the actual optimum. {\color{black}Finally, Fig.~\ref{fig:comp} confirms that the received power threshold of $-20$ dBm to $-30$ dBm required to activate practical EH receivers \cite{Lu-14-A} is met by our algorithm, even with the most conservative choice of simulation parameters (i.e., the smallest number of antennas $M=6$, a maximum distance of $9$ m from the BS to EH-ID UEs, and the smallest BS power $P=20$ dBm).}

\section{Conclusions}
We have proposed  successive second-order cone programming algorithms for
the joint design of transmit beamforming vectors and receive signal splitting factors. The objective is to maximize either the
sum EH or the energy harvested by the receiver with the least favorable channel conditions
under the UE SINR and the BS power constraints. Simulation results with practical parameters have confirmed the merits of the proposed algorithms.

\section*{Appendix: Proof of Lemma \ref{prop}}
As function $f(z)=|z|^2$ is convex in $z\in \mathbb{C}$, the so-called perspective of
$f(z)$ defined as $\bar{f}(z,y)=yf(z/y)=|z|^2/y$ is also convex in $z\in\mathbb{C}$ and $y>0$ \cite{Boyd-04-B}. This gives
\begin{align}
|z|^2/y\geq&\ds |\bar{z}|^2/\bar{y}+\la \nabla \bar{f}(\bar{z},\bar{y}),(z,y)-(\bar{z},\bar{y})\ra\nonumber\\ 
=&2\Re\{\bar{z}^Hz\}/\bar{y}-|\bar{z}|^2y/\bar{y}^2, \, \forall z, \bar{z}  \text{ and } y>0, \bar{y}>0. \label{ap1}
\end{align}

Upon rewriting
$(1-\alpha_{n_1}^2)|\mathbf{h}_{n_1}^H \mathbf{w}_{{\color{black} \eta}}|^2=|\mathbf{h}_{n_1}^H \mathbf{w}_{{\color{black} \eta}}|^2/(1-\alpha_{n_1}^2)^{-1}$
for {\color{black}$\eta \in \mathcal{N}$} and applying (\ref{ap1})  for
$z=\mathbf{h}_{n_1}^H \mathbf{w}_{{\color{black} \eta}}, \bar{z}=\mathbf{h}_{n_1}^H \mathbf{w}_{{\color{black} \eta}}^{(\kappa)}$
and
$y=1/(1-\alpha_{n_1}^2)$, $\bar{y}=1/(1-(\alpha_{n_1}^{(\kappa)})^2)$,
we have
\begin{align}
(1-\alpha_{n_1}^2)|\mathbf{h}_{n_1}^H \mathbf{w}_{{\color{black} \eta}}|^2 &\geq  \ds 2(1-(\alpha_{n_1}^{(\kappa)})^2)
        \Re\{(\mathbf{w}_{{\color{black} \eta}}^{(\kappa)})^H\mathbf{h}_{n_1}\mathbf{h}^H_{n_1}
        \mathbf{w}_{{\color{black} \eta}} \}\nonumber\\
          &\quad -  |\mathbf{h}_{n_1}^H \mathbf{w}_{{\color{black} \eta}}^{(\kappa)}|^2(1-(\alpha_{n_1}^{(\kappa)})^2)^2)/(1-\alpha_{n_1}^2).\nonumber
\end{align}
Similarly,
$(1-\alpha_{n_1}^2)\sigma_a^2\geq 2\sigma_a^2(1-(\alpha_{n_1}^{(\kappa)})^2)-\sigma_a^2(1-(\alpha_{n_1}^{(\kappa)})^2)^2/(1-\alpha_{n_1}^2).$ 
Recall that the LHS of (\ref{in1}) is $(1-\alpha_{n_1}^2)p_{n_1}(\mathbf{w})=\sum_{{\color{black} \eta}\in\mathcal{N}}(1-\alpha_{n_1}^2)| \mathbf{h}_{n_1}^H \mathbf{w}_{{\color{black} \eta}}  |^2  + (1-\alpha_{n_1}^2)\siga$. Then,
\begin{align}
&\mbox{LHS of} \ \eqref{in1}\nonumber\\
&\geq \ds\sum_{{\color{black} \eta}\in\mathcal{N}}[2(1-(\alpha_{n_1}^{(\kappa)})^2)
        \Re\{(\mathbf{w}_{{\color{black} \eta}}^{(\kappa)})^H\mathbf{h}_{n_1}\mathbf{h}^H_{n_1}
        \mathbf{w}_{{\color{black} \eta}} \}\nonumber\\
& \qquad-  |\mathbf{h}_{n_1}^H \mathbf{w}_{{\color{black} \eta}}^{(\kappa)}|^2(1-(\alpha_{n_1}^{(\kappa)})^2)^2/(1-\alpha_{n_1}^2)]\nonumber\\
&\qquad +2\sigma_a^2(1-(\alpha_{n_1}^{(\kappa)})^2)-\sigma_a^2(1-(\alpha_{n_1}^{(\kappa)})^2)^2/(1-\alpha_{n_1}^2)\nonumber\\
&=2(1-(\alpha_{n_1}^{(\kappa)})^2)[\ds  \sum_{{\color{black} \eta}\in\mathcal{N}}
        \Re\{(\mathbf{w}_{{\color{black} \eta}}^{(\kappa)})^H\mathbf{h}_{n_1}\mathbf{h}^H_{n_1}
        \mathbf{w}_{{\color{black} \eta}} \}+\sigma_a^2]\nonumber\\
&\qquad-\ds p_{n_1}(\mathbf{w}^{(\kappa)})(1-(\alpha_{n_1}^{(\kappa)})^2)^2/(1-\alpha_{n_1}^2),
\end{align}
showing the inequality in (\ref{in1}). The equality in (\ref{in2}) is obvious.


\end{document}